\documentclass[usenatbib]{mn2e} %MN-12-1030-MJ submitted May 5 2012 - resubmitted July 7
\usepackage{natbib}
\usepackage{epsfig}

\catcode`\@=11
\def\gta{\ifmmode{\,\mathrel{\mathpalette\@versim>\,}}
    \else{$\,\mathrel{\mathpalette\@versim>}\,$}\fi}
\def\lta{\ifmmode{\,\mathrel{\mathpalette\@versim<\,}}
    \else{$\,\mathrel{\mathpalette\@versim<}\,$}\fi}
\def\@versim#1#2{\lower 2.9truept \vbox{\baselineskip 0pt \lineskip
    0.5truept \ialign{$\m@th#1\hfil##\hfil$\crcr#2\crcr\sim\crcr}}}
\catcode`\@=12  
\renewcommand{\[}{\begin{equation}}
\renewcommand{\]}{\end{equation}}
{\newif\ifnotend
\notendtrue
\def\veclist{ABCDEFGHIJKLMNOPQRSTUVWXYZabcdefghijklmnopqrstuvwxyz.}
\def\top#1#2.{#1}
\def\tail#1#2.{#2.}
\loop\expandafter\xdef\csname v\expandafter\top\veclist\endcsname%
{{\noexpand\bf\expandafter\top\veclist}}
\edef\veclist{\expandafter\tail\veclist}
\if\veclist.\notendfalse\fi\ifnotend\repeat}
{\newif\ifnotend
\notendtrue
\def\veclist{ABCDEFGHIJKLMNOPQRSTUVWXYZ.}
\def\top#1#2.{#1}
\def\tail#1#2.{#2.}
\loop\expandafter\xdef\csname c\expandafter\top\veclist\endcsname%
{{\noexpand\cal\expandafter\top\veclist}}
\edef\veclist{\expandafter\tail\veclist}
\if\veclist.\notendfalse\fi\ifnotend\repeat}

\def\df{{\sc df}}

\def\fracj#1#2{{\textstyle{#1\over#2}}}

\def\Rc{R_{\rm c}}\def\vc{v_{\rm c}}

\def\kms{\,{\rm km}\,{\rm s}^{-1}}

\def\Gyr{\,{\rm Gyr}}
 
\def\pc{\,{\rm pc}}
\def\kpc{\,{\rm kpc}}

\def\e{{\rm e}}
\def\d{{\rm d}}
\def\msun{\,{\rm M}_\odot}

\def\vpbar{\langle{v}_\phi\rangle}
\def\figref#1{Fig.~\ref{#1}}
\newcommand{\beq}{\begin{equation}}
\newcommand{\eeq}{\end{equation}}

\title[More dynamical models of our Galaxy]
{More dynamical models of our Galaxy}

\author[James Binney]{James  Binney\thanks{E-mail:
binney@thphys.ox.ac.uk}\\
Rudolf Peierls Centre for Theoretical Physics, Keble Road, Oxford OX1 3NP, UK\\
}

\begin{document}

\date{Draft, July 5, 2012}

\pagerange{\pageref{firstpage}--\pageref{lastpage}} \pubyear{2012}

\maketitle

\label{firstpage}

\begin{abstract}
A companion paper presents an algorithm for estimating the actions of orbits
in axisymmetric potentials. This algorithm is fast enough for it to be
feasible to fit automatically a parametrised distribution function to
observational data for the solar neighbourhood. We explore the predictive
power of these models and the extent to which global models are constrained
by data confined to the solar cylinder.  We adopt a gravitational potential
that is generated by three discs (gas and both thin and thick stellar discs),
a bulge and a dark halo, and fit the thin-disc component of the distribution
function to the solar-neighbourhood velocity distribution from the
Geneva-Copenhagen Survey.  We find that the disc's vertical density profile
is in good agreement with data at $z\lta500\pc$.  The thick-disc component of
the distribution function is then used to extend the fit to data from Gilmore
\& Reid (1983) for $z\lta2.5\kpc$. The resulting model predicts excellent
fits to the profile of the vertical velocity dispersion $\sigma_z(z)$ from
the RAVE survey and to the distribution of $v_\phi$ velocity components at
$|z|\sim1\kpc$ from the SDSS survey. The ability of this model to predict
successfully data that was not used in the fitting process suggests that the
adopted gravitational potential (which is close to a maximum-disc potential)
is close to the true one.  We show that if another plausible potential is
used, the predicted values of $\sigma_z$ are too large. The models imply that
in contrast to the thin disc, the thick disc has to be hotter vertically than
radially, a prediction that it will be possible to test in the near future.
When the model parameters are adjusted in an unconstrained manner, there is a
tendency to produce models that predict unexpected radial variations in
quantities such as scale height. This finding suggests that to constrain
these models adequately one needs data that extends significantly beyond the
solar cylinder. The models presented in this paper might prove useful to the
interpretation of data for external galaxies that has been taken with an
integral field unit.
\end{abstract}

\begin{keywords}
galaxies: kinematics and dynamics
- The Galaxy: disc - solar neighbourhood
\end{keywords} 

\section{Introduction}

Large-scale surveys of our Galaxy are underway and in 2013 the European Space
Agency will launch a satellite, Gaia, that is tasked with determining
astrometry and photometry of unprecedented precision for a billion stars and
gathering the spectra of a hundred million stars. The large outlays required
to gather these data have been motivated by the expectation that we will be
able to infer from the data not only the distribution of the Galaxy's dark
matter, but also quite detailed knowledge of the manner of its formation and
its evolutionary history.  Dynamical models of the Galaxy will be central to
achieving these goals. 

The simplest plausible dynamical models approximate the Galaxy by an
axisymmetric body and exploit Jeans' theorem to make the distribution
function (\df) dependent on just three isolating integrals. There are
substantial advantages in identifying these integrals with the actions $J_r$,
which quantifies a star's radial oscillations, $J_z$, which quantifies
oscillations perpendicular to the Galaxy's equatorial plane, and $L_z$, the
component of angular momentum about the assumed symmetry axis.

It turns out that good fits to the available observational data can be
obtained with models whose \df s are simple analytic functions of the actions
\citep[][hereafter B10]{B10}. Given such a \df, the calculation of
predictions that can be compared with data is greatly facilitated if it is
easy to determine the actions $\vJ$ of a given phase-space point $(\vx,\vv)$.
Analytic expressions for $\vJ(\vx,\vv)$ are not available for any realistic
Galactic potential and one has to have recourse to approximate and numerical
methods. In B10 and \cite{BinneyM} the observable properties of models were
obtained from the `adiabatic approximation' for actions. In a companion paper
we show that an algorithm based on the proximity of Galactic potentials to
St\"ackel potentials yields more accurate estimates of actions for a wider
class of orbits. Moreover, this algorithm can be implemented in a
sufficiently streamlined way that the observables of $\sim50$ models can be
estimated per hour on a laptop, and it becomes relatively straightforward to
search the space of possible \df s automatically rather than by hand, and
heavily influenced by prior prejudice, as was done in B10.

The purpose of this paper is to present results obtained by such automatic
searches. Our aim is to explore the extent to which the global structure of
the Galaxy can be pinned down by restricted sets of data when we impose a
particular functional form for the \df. The data we consider are restricted
to the solar cylinder and for the most part quite old, so the models we
obtain are far from definitive. Surveys now in hand will shortly yield data
with much better statistics that extend significantly beyond the solar
cylinder, so now is not the time to seek definitive results. Rather it is the
moment to explore possibilities and connections between different types of
data, and these are the tasks addressed in this paper.

Section \ref{sec:pots} we describes the adopted potentials and Section
\ref{sec:DFs} gives the functional forms of the adopted distribution
functions. Section \ref{sec:models} shows fits obtained to observational data
using two potentials, which differ in the assumed values of the distance
$R_0$ to the Galactic centre and the local circular speed $\Theta_0$. Section
\ref{sec:discuss} demonstrates that the thick disc has to be hotter
vertically than radially, and addresses a variety of issues that are raised by
the models.  Section \ref{sec:conclude} sums up and considers what
should be done next in relation to both surveys of our Galaxy and of external
galaxies.  An Appendix explains how we evaluate the multiple integrals over
velocity that extract observables from the \df.

\begin{table*}
\begin{center}
\caption{Parameters of the potentials}\label{tab:pots}
\begin{tabular}{lccc|ccc}
\hline
&\multispan3{\hfil Potential I\hfil}&\multispan3{\hfil Potential II\hfil}\\
Disc&Thin&Thick&Gas&Thin&Thick&Gas\\
\hline
$\Sigma_0[\!\msun\kpc^{-2}]$&1.02e9&1.14e6&7.30e7
					&7.68e8&2.01e8&1.16e8\\
$R_\d[\!\kpc]$&2.4&2.4&4.8              &2.64&2.97&5.28\\
$z_\d[\!\kpc]$&0.36&1&0.04		&0.3&0.9&0.04\\
$R_{\rm h}[\!\kpc]$&0&0&4.0             &0&0&4\\
\hline
Spheroid&Dark&Stellar&&Dark&Stellar\\
\hline
$\rho_0[\!\msun\kpc^{-3}]$&1.26e9&7.56e8&
					&1.32e7&9.49e10\\
$q$&0.8&0.6&				&1&0.5\\
$\gamma$&-2&1.8&                        &1&0\\
$\beta$&2.21&1.8&                       &3&1.8\\
$r_0[\!\kpc]$&1.09&1&                   &16.47&0.075\\
$r_{\rm cut}[\!\kpc]$&1000&1.9&         &100000&2.1\\
\hline
\end{tabular}
\end{center}
\end{table*}

\section{Gravitational potentials}\label{sec:pots}

We have worked with two gravitational potentials of the type presented by
\cite{DehnenB}. Each potential is generated by three superposed discs: one
representing the gas layer, one the thin disc and one representing the thick
disc. The density of each disc is given by
\[
\rho(R,z)={\Sigma_0\over 2z_\d}\exp\left[-\left({R_{\rm h}\over R}+{R\over
R_\d}+{|z|\over z_\d}\right)\right],
\]
 where a non-zero value of $R_{\rm h}$ generates a central depression in an
otherwise double-exponential disc. For each disc Table~\ref{tab:pots} gives
the values taken by the parameters that appear in this formula. Spheroids
representing the bulge and the dark halo also contribute to the potentials.
The density of each spheroid is given by
\[
\rho(R,z)={\rho_0\over m^\gamma(1+m)^{\beta-\gamma}}\exp\left[-(mr_0/r_{\rm
cut})^2\right],
\]
 where
\[
m(R,z)\equiv\sqrt{(R/r_0)^2+(z/qr_0)^2}.
\]
 Table~\ref{tab:pots} gives the values of the parameters for each spheroid.
Potential I assumes $R_0=8\kpc$ and differs from Model 2 of \cite{DehnenB}
only in having the scale height of the thin disc increased from $z_\d=180\pc$
to $z_\d=350\pc$ and having the mass of the thin disc adjusted to increase
the local circular speed from $\Theta_0=217\kms$ to $\Theta_0=220\kms$. This
potential has a fairly short disc scale-length, so it is nearly a
maximal-disc model.  Potential II assumes $R_0=8.37\kpc$. It has been chosen
to satisfy the constraints listed in \cite{McMillan11} and gives
$\Theta_0=241\kms$.

\begin{figure}
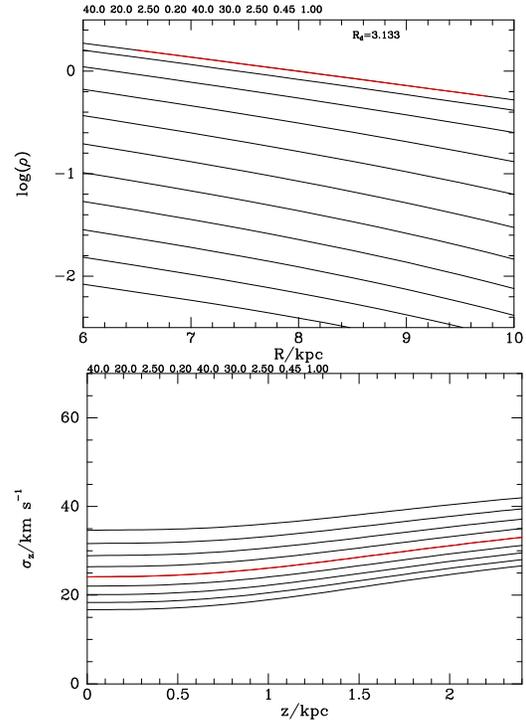

\centerline{\epsfig{file=c5/m2b_iso_q45_rhoR.ps,width=.8\hsize}}
\centerline{\epsfig{file=c5/m2b_iso_q45_sigz.ps,width=.8\hsize}}
\caption{Properties of a quasi-isothermal component with
$(\sigma_{r0},\sigma_{z0})=(40,20)\kms$, $R_\d=2.5\kpc$ and $q=0.45$. 
Top: density as a function of radius at heights $z$ that increase by
$0.25\kpc$ from $z=0$ at the top. Bottom: vertical velocity dispersion as a
function of $z$ at radii that from top to bottom increase from $6\kpc$ to
$10\kpc$. In the top panel two points on the density profile in the plane are
joined by a straight red line, and this line is an exponential with
scalelength $R_\d=3.13\kpc$.}\label{fig:iso}
\end{figure}

\section{Distribution functions}\label{sec:DFs}

Our \df s are built up out of ``quasi-isothermal'' components. The \df\ of such a
component is
 \[\label{eq:qi}
f(J_r,J_z,L_z)=f_{\sigma_r}(J_r,L_z)f_{\sigma_z}(J_z,L_z),
\] 
 where $f_{\sigma_r}$ and $f_{\sigma_z}$ are defined to be
 \[\label{planeDF}
f_{\sigma_r}(J_r,L_z)\equiv{\Omega\Sigma\over\pi\sigma_r^2\kappa}
[1+\tanh(L_z/L_0)]\e^{-\kappa J_r/\sigma_r^2}
\]
 and
 \[\label{basicvert}
f_{\sigma_z}(J_z,L_z)\equiv{\nu\over2\pi\sigma_z^2}\,\e^{-\nu J_z/\sigma_z^2}.
\]
 Here $\Omega(L_z)$, $\kappa(L_z)$ and $\nu(L_z)$ are the circular, radial
and vertical epicycle frequencies respectively, while
\[\label{eq:defsSigma}
\Sigma(L_z)=\Sigma_0\e^{-\Rc/R_\d}
\]
 is the approximate surface density of the disc, with $\Rc(L_z)$ the radius
of the circular orbit with angular momentum $L_z$. The functions
$\sigma_r(L_z)$ and $\sigma_z(L_z)$ control the radial and vertical velocity
dispersions in the disc and are approximately equal to them at $\Rc$. Given
that the scale heights of galactic discs do not vary strongly with radius
\citep{vdKSearle}, these quantities must increase inwards. We adopt the
following dependence on $L_z$:
 \begin{eqnarray}
\sigma_r(L_z)&=&\sigma_{r0}\,\e^{q(R_0-\Rc)/R_\d}\cr
\sigma_z(L_z)&=&\sigma_{z0}\,\e^{q(R_0-\Rc)/R_\d},
\end{eqnarray}
 which imply that the radial scale-length on which the velocity dispersions
decline is $R_\d/q$. Our expectation is that $q\sim0.5$.

 In equation (\ref{planeDF}) the factor containing tanh serves to eliminate
retrograde stars; the value of $L_0$ controls the radius within which
significant numbers of retrograde stars are found, and should be no larger
than the circular angular momentum at the half-light radius of the bulge.
Provided this condition is satisfied, the results for the solar cylinder
presented here are essentially independent of $L_0$.

\figref{fig:iso} shows an example of a quasi-isothermal component. The upper
panel shows that away from
the plane its density is quite close to exponential in both $R$ and $z$ and
the lower panel shows that the vertical velocity dispersion is independent of $z$ for $z\lta500\pc$.

\begin{figure*}
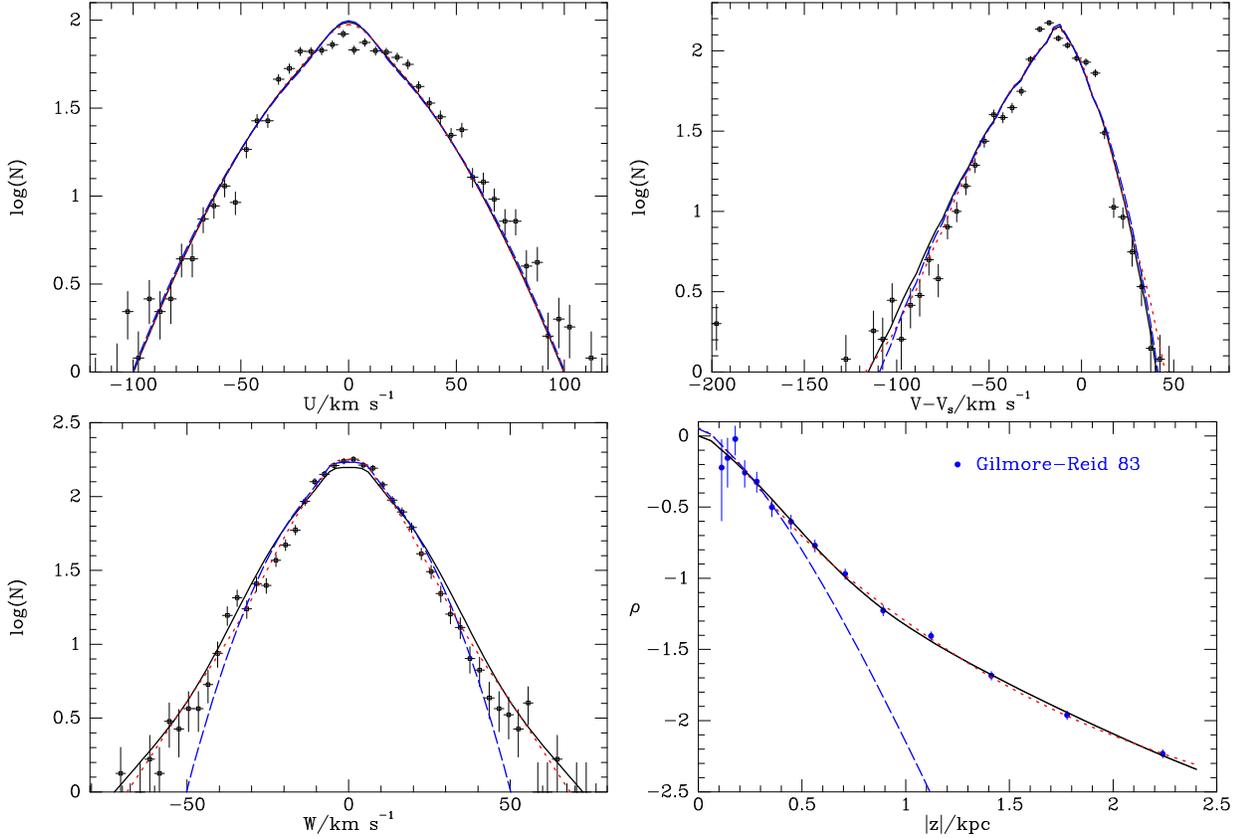

\centerline{\epsfig{file=c5/m2e_U.ps,width=.45\hsize}\quad
\epsfig{file=c5/m2e_V.ps,width=.45\hsize}}
\centerline{\epsfig{file=c5/m2e_W.ps,width=.45\hsize}\quad
\epsfig{file=c5/m2e_rho.ps,width=.45\hsize}}
 \caption{Fitting models in Potential I.  The blue dashed curves show
the result of choosing the parameters of the thin-disc \df\ to optimise the
fits of this model with no thick disc to the GCS velocity distributions of
local stars shown in the first three panels. The full curves show the results
obtained when a thick disc is included and the Gilmore-Reid points for the
density shown in the bottom-right panel are included in the data to be
fitted, without adjusting the previously-determined thin-disc \df. The red
dotted curves show the fits obtained when the parameters of both discs
are simultaneously adjusted to optimise the fits to the GCS histograms and
the Gilmore-Reid points. The parameters of the \df s responsible for the blue
dashed, full and red-dashed curves are respectively listed in columns (a) to
(c) of Table~\ref{tab:fitI}, respectively.}\label{fig:m2eA}
\end{figure*}

\begin{figure*}
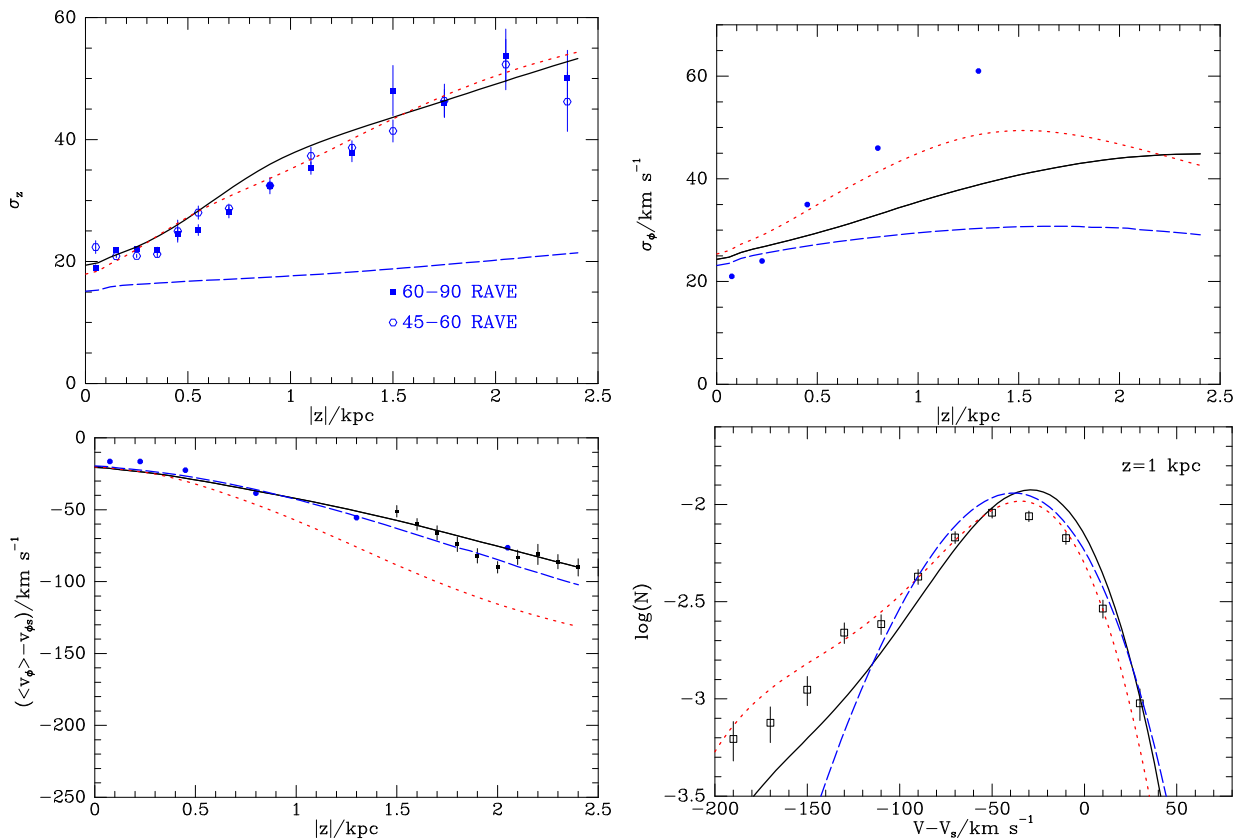

\centerline{\epsfig{file=c5/m2e_sz.ps,width=.45\hsize}\quad
\epsfig{file=c5/m2e_sp.ps,width=.45\hsize}}
\centerline{\epsfig{file=c5/m2e_vp.ps,width=.45\hsize}\quad
\epsfig{file=c5/m2e_vp1.ps,width=.45\hsize}}
 \caption{Prediction of the models fitted to to data as described in
Fig.~\ref{fig:m2eA}. The blue dashed curves are predictions for the
model that has no thick disc, the full black curves are for the model
obtained by adding a thick disc without adjusting the thin disc, and the red
dotted curves show the model obtained by adjusting simultaneously both
discs.  Blue data points are from Burnett (2010) and black ones from
Moni-Bidin et al.\ (2012). The model predictions for $\langle v_\phi\rangle$ at
$z=1\kpc$ have been convolved with a Gaussian of dispersion $19\kms$, which
is the estimated error of the SDSS data shown from Ivezic et al.\
(2008).}\label{fig:m2eB}
\end{figure*}

The \df\ defined by equation (\ref{planeDF}) is the planar
``pseudo-isothermal'' \df\ of B10, while that defined by equation
(\ref{basicvert}) differs from the vertical ``pseudo-isothermal'' in B10 only
in the replacement in the exponential of the vertical frequency
$\Omega_z(\vJ)$ by the vertical epicycle frequency $\nu(L_z)$. This
replacement is expedient because at large radii $r$, where the potential
becomes quite nearly spherical, $J_z\to L-|L_z|$ so for an orbit with $J_r=0$
$J_z\sim\vc r$, while $\Omega_z\to\Omega\sim \vc/r$, so $\Omega_zJ_z\to$ a
constant.  Consequently, B10's \df\ tends to a constant at large $J_z$ and
fixed $L_z$, which is inappropriate.

The functions $f_{\sigma_i}$ satisfy the normalisation conditions
 \begin{eqnarray}
\int_0^\infty\d J_r\,f_{\sigma_r}&
=&{\Omega\Sigma\over\pi\kappa^2}[1+\tanh(L_z/L_0)]\cr
\int_0^\infty\d J_z\,f_{\sigma_z}&=&{1\over2\pi},
\end{eqnarray}
 so 
\[
g(L_z)\equiv\int\d J_r\int\d J_z\,f(J_r,J_z,L_z),
\]
 which is the number of stars per unit angular momentum, decreases as
$\Sigma(L_z)/\kappa(L_z)$, so roughly exponentially.

We take the \df\ of the thick disc to be a pseudo-isothermal. The thin
disc is treated as a superposition of the cohorts of stars that have age
$\tau$ for ages that vary from zero up to the age $\tau_{\rm
max}\simeq10\Gyr$ of the thin disc. We take the \df\ of each such cohort to
be a pseudo-isothermal with velocity-dispersion parameters $\sigma_{r}$ and
$\sigma_{z}$ that depend on age as well as on $L_z$.  Specifically, from 
\cite{AumerB}  we adopt
 \begin{eqnarray}\label{sigofLtau}
\sigma_r(L_z,\tau)&=&\sigma_{r0}\left({\tau+\tau_1\over\tau_{\rm
m}+\tau_1}\right)^\beta\e^{q(R_0-\Rc)/R_\d}\nonumber\\
\sigma_z(L_z,\tau)
&=&\sigma_{z0}\left({\tau+\tau_1\over\tau_{\rm m}+\tau_1}\right)^\beta
\e^{q(R_0-\Rc)/R_\d}.
\end{eqnarray}
 Here $\sigma_{z0}$ is the approximate vertical velocity dispersion of local
stars at age $\tau_{\rm m}\simeq10\Gyr$, $\tau_1$ sets velocity dispersion at
birth, and $\beta\simeq0.33$ is an index that determines how the velocity
dispersions  grow
with age.  We further assume that the star-formation rate in the thin disc
has decreased exponentially with time, with characteristic timescale $t_0$,
so the thin-disc \df\ is
 \[\label{thinDF}
f_{\rm thn}(J_r,J_z,L_z)={\int_0^{\tau_{\rm m}}\d\tau\,\e^{\tau/t_0}
f_{\sigma_r}(J_r,L_z)f_{\sigma_z}(J_z,L_z)
\over t_0(\e^{\tau_{\rm m}/t_0}-1)},
\]
 where $\sigma_r$ and $\sigma_z$ depend on $L_z$ and $\tau$ through equation
(\ref{sigofLtau}). We set the normalising constant $\Sigma_0$ that appears in
equation (\ref{eq:defsSigma}) to be the same for both discs and use for the
complete \df
 \[
f(J_r,J_z,L_z)=f_{\rm thn}(J_r,J_z,L_z)+Ff_{\rm thk}(J_r,J_z,L_z),
\]
 where $F$ is a parameter that controls the fraction of stars that belong to
the thick disc.

The \df s of the thin and thick discs each involve four important parameters,
$\sigma_{r0}$, $\sigma_{z0}$, $R_\d$ and $q$. The \df\ of the thin disc
involves three further parameters, $\tau_1$, $\tau_{\rm m}$ and $\beta$, but
we shall not explore the impact of changing these here because we do not
consider data that permit discrimination between stars of different ages.
Therefore following \cite{AumerB} we adopt throughout $\tau_1=0.01\Gyr$,
$\tau_{\rm m}=10\Gyr$ and $\beta=0.33$.

We have used the {\it amoeba} routine of \cite{Pressetal} to adjust nine
parameters of the overall \df: $\sigma_{r0}$, $\sigma_{z0}$, $R_\d$, and $q$
for the thick and the thin discs plus the relative weight $F$ of
the thick and thin discs. 

\section{models}\label{sec:models}

The procedure generally adopted was to have {\it amoeba} fit the \df\ to the
$U$, $V$ and $W$ histograms for solar-neighbourhood stars from the
Geneva-Copenhagen survey \citep[][hereafter GCS]{Nordstrom04,HolmbergNA}
using only a thin disc, and then to add a thick disc to the \df\ and use its
parameters to secure a fit to vertical density profile of F dwarfs inferred
by \cite{GilmoreR}. In a final step {\it amoeba} adjusted all nine parameters
of the \df\ simultaneously to polish the fit to the GCS histograms and the
Gilmore-Reid points.  

The histograms fitted at each stage were compiled using all GCS stars closer
than $150\pc$ with a probability of a constant line-of-sight velocity $>0.3$.
The $U$ and $W$ components have been shifted to the Local Standard of Rest
frame using $U_\odot=11.1\kms$ and $W_\odot=7.25\kms$ from
\cite{SchoenrichBD}. The $V$ components were heliocentric.

In the second and third stages of fitting, the quantity to be minimised is
 \[\label{eq:chisq}
\chi^2=\fracj12(\chi^2_U+\chi^2_V+\chi^2_W)+3\chi^2_\rho,
\]
 where each component, $\chi_U^2$ etc., is the mean-square ratio of the
difference between model and data divided by the formal observational error,
and the sum of $U,V,W$ terms is what was minimised in the first stage of
fitting. The relative weighting of the velocity and density data is an
arbitrary choice designed to ensure that the relatively small number of
density data are taken seriously. The iterations stop when the fractional
variation of $\chi^2$ across the simplex is $<10^{-4}$.

\subsection{Fits in Potential I}

\figref{fig:m2eA} shows the fits obtained in Potential I. All three \df s
provides similar fits to the histograms of $U$ and $V$, but the \df\ without
a thick disc (blue dashed lines) falls below the data at large $|W|$ and
$z\gta500\pc$ as is to be expected. The other two \df s provide excellent
fits to the data apart from minor discrepancies within the cores of the $U$
and $V$ distributions. These discrepancies probably reflect the impact on the
GCS histograms of non-equilibrium structure that lies beyond the scope of the
present models.  In particular, asymmetries in the observed distributions of
$U$ and $W$ components cannot be reproduced by an equilibrium model.  The
bottom two panels of Fig.~\ref{fig:m2eA} provides two indications that the
Galaxy's true potential does not differ greatly from Potential I. First even
though the thin-disc-only \df\ was fitted only to the velocity data, it does
provide a reasonable fit to the Gilmore-Reid points in the region
$z\lta500\pc$ dominated by the thin disc. Second, the other two \df s can
simultaneously fit both the $W$ distribution and the Gilmore-Reid points -- in
an erroneous potential it should be possible to fit either of  these datasets
but not both simultaneously.

Fig.~\ref{fig:m2eB} compares the predictions of these \df s with data that
were not used in the fitting process. Each \df\ is shown by the same line
type as in Fig.~\ref{fig:m2eA}. The blue data points come from Burnett
(2010), black ones come from \cite{Moni-B12} and the open points in the
bottom-right panel for the $V$ distribution at $|z|=1\kpc$ come from
\cite{Ivezic08}. The predictions of the \df s shown in the bottom-right
panel have been convolved with a Gaussian distribution of dispersion
$19\kms$, the observational uncertainty reported by \cite{Ivezic08}.
Burnett's blue data points are the fruit of a preliminary analysis of
$\sim200\,000$ stars in the RAVE survey, roughly half dwarfs and half giants.
Error bars are not available for the measures of $\langle v_\phi\rangle$ and
$\sigma_\phi$. The black data points from Moni-Bidin et al.\ are obtained
from a sample of 412 red giants seen near the south Galactic pole. They are
shown with the errors given by \cite{Moni-B12} but these are significantly
too small \citep{Sanders12}. The open data points of Ivezic et al.\ relate to
a very large sample of dwarf stars in the Sloan Digital Sky Survey.

Rather than plotting $\langle v_\phi\rangle$ we plot this less the value in
the model of the Sun's azimuthal velocity $v_{\phi\rm s}=\Theta_0+V_{\rm s}$,
and we compare with heliocentric values of $v_\phi$. This comparison is to
first order insensitive to the uncertain peculiar azimuthal velocity of the
Sun, $V_\odot\simeq11.5\kms$ \citep{SchoenrichBD}.

As the points from Ivezic et al.\ illustrate, the distribution in $v_\phi$ is
expected to be very skew and cannot be accurately characterised by a mean and
a dispersion, especially far from the plane. Moreover, our \df s are designed
to provide only disc stars, and far from the plane halo stars will make
non-negligible contributions to the velocity distributions, especially at
small $v_\phi$. So rather than comparing the predicted and measured values of
$\langle v_\phi\rangle$ and $\sigma_\phi$ at various heights, we should judge
a model on how well it reproduces the complete $v_\phi$ distribution at
several values of $z$, as is done in the bottom-right panel of
\figref{fig:m2eB}.

In the top left panel of Fig.~\ref{fig:m2eB} we see that, as expected, the
thin-disc-only model predicts a rather constant value of $\sigma_z$ that lies
below the data at all $z$. By contrast both models with thick discs fit the
data to an extent that is remarkable given that the data played no part in
choosing these models. The ability of these models to predict the run of
$\sigma_z(z)$ is a further indication that Potential I does not differ
greatly from the Galaxy's potential. 

The bottom-right panel of Fig.~\ref{fig:m2eB} shows that the red dotted line
provides a good fit to the $v_\phi$ distribution at $z\simeq1\kpc$ from
\cite{Ivezic08} aside from predicting slightly too many stars at
$v_\phi-V_{\rm s}\lta-100\kms$. This defect is unfortunate because on account
of its neglect of the stellar halo, the model should {\it undershoot\/} the
data in this region.  The lower left panel of Fig.~\ref{fig:m2eB} shows that
in this model $\langle v_\phi\rangle$ falls too rapidly with $|z|$, a result
consistent with the excess of stars at $v_\phi-V_{\rm s}<-100\kms$ in the
lower-right panel. The upper-right panel suggests that in all three models
$\sigma_\phi$ rises too gradually with $|z|$. However, this suggestion is
contradicted by the lower-right panel, which
implies that at $z=1\kpc$ the value of $\sigma_\phi$ for the model shown by the
red dotted curve {\it exceeds} that in the Galaxy. Indeed data points for
$\sigma_\phi$ from red-clump stars in RAVE prove to lie systematically below
Burnett's values (Williams et al.\ in preparation), so it seems likely that
the data points in the upper-right panel of \figref{fig:m2eB} are biased to
high values. 

Overall, we conclude that although the \df\ in which all parameters have been
simultaneously adjusted (red dotted lines) gives a remarkably good account of
data that was not involved in its choice, a more perfect account of the data
would be given by a \df\ that is intermediate between this \df\ and the one
determined by fixing the thin and thick discs independently (full curves).

One finds, not surprisingly, that models with higher
$\sigma_\phi$ tend to have lower $\langle v_\phi\rangle$, and vice versa.

Columns (a) -- (c) of Table~\ref{tab:fitI} give the parameters of the \df s
of the models shown in Figs~\ref{fig:m2eA} and \ref{fig:m2eB}. In column (a)
we see that there is nothing remarkable about the parameters of the thin disc
initially chosen. The bottom half of column (b) shows that the thick disc
that was selected to complement this thin disc has a remarkably small value
of $\sigma_{r0}$ ($25.8\kms$), and a remarkably large normalisation
($F=0.772$), which implies that $\sim43$ per cent of all stars are in the
thick disc. Column (c) shows that an effect of simultaneously adjusting all
nine parameters of the \df\ is to weaken the radial gradient of $\sigma_r$ in
the thin disc ($q=0.29\to q=0.14$) and to increase the gradient of $\sigma_r$
in the thick disc ($q=0.52\to q=0.71$). Another surprising effect is to
increase the normalisation of the thick disc to $F=1.4$, so now 58 per cent
of all stars lie in the thick disc.

When considering multi-parameter models such as these one should ask how
unique a given fit to data really is.  An indication is given by column (d)
of Table~\ref{tab:fitI}, which gives the parameters of the \df\ obtained by
dispensing with a preliminary fit of the thin disc to the GCS data and from
the outset simultaneously adjusting all nine parameters to optimise the fit
to the GCS velocity histograms and the Gilmore-Reid density points. This \df\
provides a fit to the given data which is barely distinguishable from that
provided by the \df\ of column (c) (red dotted lines), and very similar
predictions to those plotted in \figref{fig:m2eB}; the only significant
difference is that with column (d) at $z=1\kpc$ $\langle v_\phi\rangle$ is
predicted to be $\sim7\kms$ higher and $\sigma_\phi$ a similar amount lower
than with column (c). There are however quite significant differences in the
\df s: the thin-disc scale length is $2.80\kpc$ in column (c) and $2.17\kpc$
in column (d), and in the thin disc of column (d) the radial gradient in
$\sigma_r$ virtually vanishes. Conversely, the thick-disc scale length is
$2.5\kpc$ in column (c) and $3.66\kpc$ in column (d) while the already steep
radial gradient of $\sigma_r$ in the thick disc has steepened to $q=1.07$ in
column (d) from $q=0.705$ in column (c). Notice that increases in $R_\d$ and
$q$ tend to compensate, because they tend to hold constant the scale length
$R_\d/q$ on which $\sigma_r$ decreases with $R$.  Experience shows that when
tasked with fitting any data for the solar cylinder {\it
amoeba\/} tends to choose thick discs which have large values of both $R_\d$
and $q$. One suspects that such models are not very physical and would
be excluded by observational data from outside the solar cylinder.

\begin{table}
\caption{Parameters of the \df\ chosen by {\it amoeba} for Potential I.
Column (a) shows the thin-disc \df\ chosen to optimise the fits to just the
GCS velocity distributions.  Column (b) gives the parameters obtained when we
add both a thick disc and data for $\rho(z)$. Column (c) shows the \df\
chosen when {\it amoeba} is given the opportunity to adjust all parameters
simultaneously, starting with the \df\ of column (b). In Figs.~\ref{fig:m2eA}
and \ref{fig:m2eB} the \df\ of column (a) gives rise to the blue dashed
curves, that of column (b) to the full curves, and that of column (c) to the
red dotted curves.
Column (d) shows the
result of optimising the complete \df\ in a single step, using both the GCS
data and the $\rho(z)$ from the outset. The parameters listed in columns (c)
and (d) yield very similar predictions for all
observables. The \df\ specified by Column (e) was chosen by fixing the
parameters of the thin disc at plausible values and then adjusting the
thick-disc parameters to optimise the fit to 
$\rho(z)$ and the wings of the GCS histograms for $U$ and $W$.
\figref{fig:m2eB_2} shows that this \df\ conflicts with constraints on the
$v_\phi$ distribution, especially away from the plane.}\label{tab:fitI}

\begin{center}
\begin{tabular}{llccccc}
\hline
&&(a)&(b)&(c)&(d)&(e)\\
\hline
Thin&$\sigma_{r0}$&40.1&40.1&42.2&42.3&30\\
&$\sigma_{z0}$    &25.6&25.6&19.5&20.3&20\\
&$R_\d$           &2.58&2.58&2.80&2.17&2.5\\
&$q$              &0.289&0.289&0.142&.040&0.450\\
\hline
Thick&$\sigma_{r0}$&-&25.8&25.2&26.3&39.6\\
&$\sigma_{z0}$     &-&45.0&32.7&34.0&30.4\\
&$R_\d$            &-&2.11&2.50&3.66&2.28\\
&$q$               &-&0.522&0.705&1.068&0.524\\
&$F$     &0&0.772&1.424&0.224&0.989\\
\hline
$\chi^2$&&16.8&9.40&7.61&7.44&4.51\\
\hline
\end{tabular}
\end{center}
\end{table}

\begin{figure*}
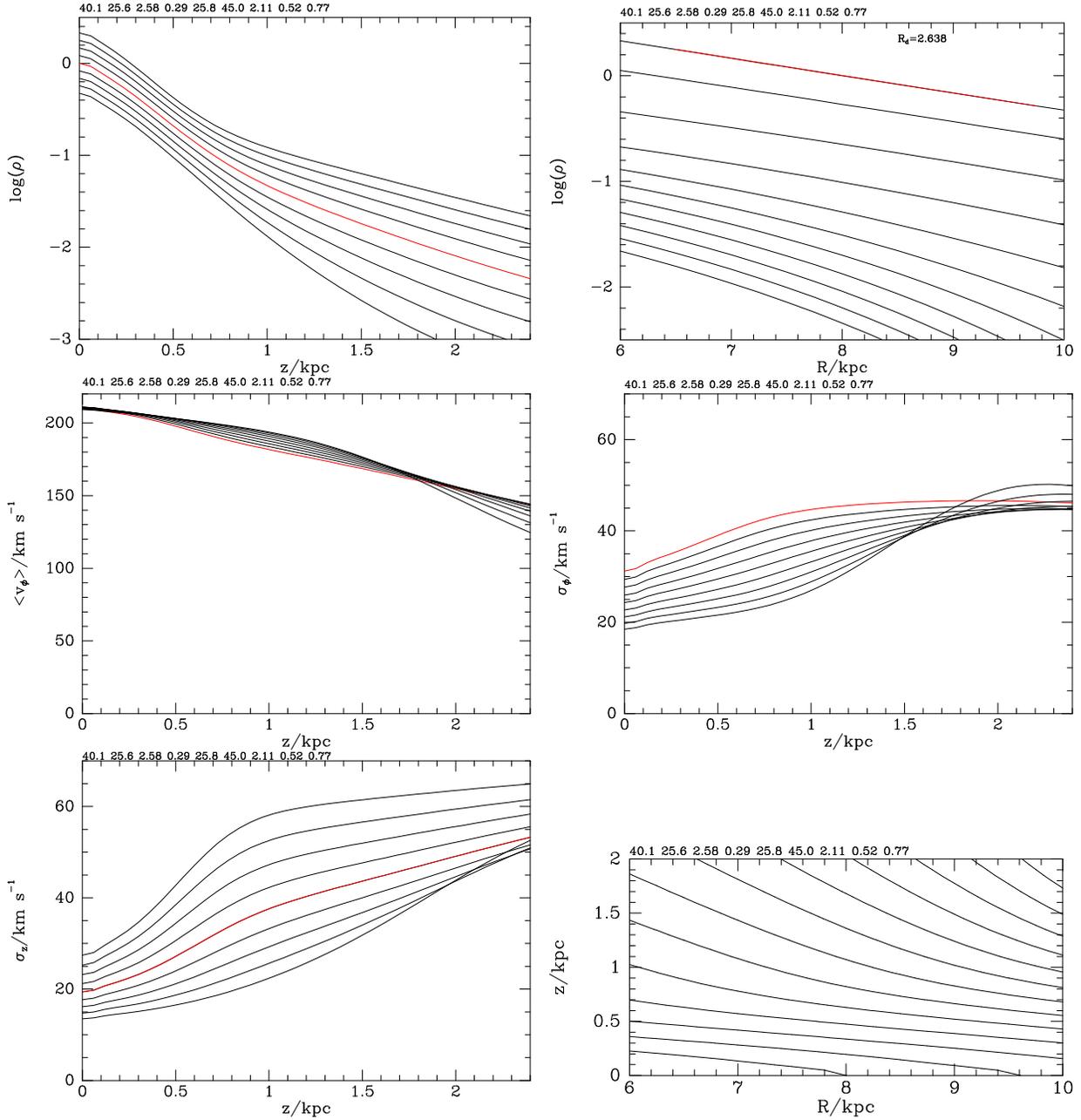

\centerline{\epsfig{file=c5/m2e_rhoz.ps,width=.45\hsize}\quad
\epsfig{file=c5/m2e_rhoR.ps,width=.45\hsize}}
\centerline{\epsfig{file=c5/m2e_vpz.ps,width=.45\hsize}\quad
\epsfig{file=c5/m2e_spz.ps,width=.45\hsize}}
\centerline{\epsfig{file=c5/m2e_sigz.ps,width=.45\hsize}\quad
\epsfig{file=c5/m2e_rhoRz.ps,width=.45\hsize}}
 \caption{Global properties of the model generated by the \df\ of Column (b)
in Table~\ref{tab:fitI} in Potential I. In the top left panel the curves
show the density at constant $R$ with $R$ increasing from 6 to $10\kpc$ in
steps of $0.5\kpc$ from top to bottom (the curve for $R=8\kpc$ is shown red),
while the top right panel shows $\rho$ at fixed $|z|$, with $|z|$ increasing
by $0.24\kpc$ from top to bottom. In the middle and bottom-left panels the
curves are again for fixed values of $R$ from $6$ to $10\kpc$, but now with
the curve for $R=6\kpc$ shown red. The bottom-right panel shows contours of
constant density in the $(R,z)$ plane.}\label{fig:global}
\end{figure*}

\subsubsection{Large-scale structure predicted by the best DF}

It is interesting to investigate the large-scale morphology of the disc
produced by the \df\ of column (b) of Table~\ref{tab:fitI} since, as we
have seen, this disc is consistent with most of the available data, which is
essentially local in character.  The upper panels of \figref{fig:global} show
how $\rho(R,z)$ depends on $z$ at fixed $R$ (left) and on radius at fixed
$|z|$ (right). 

\begin{figure*}
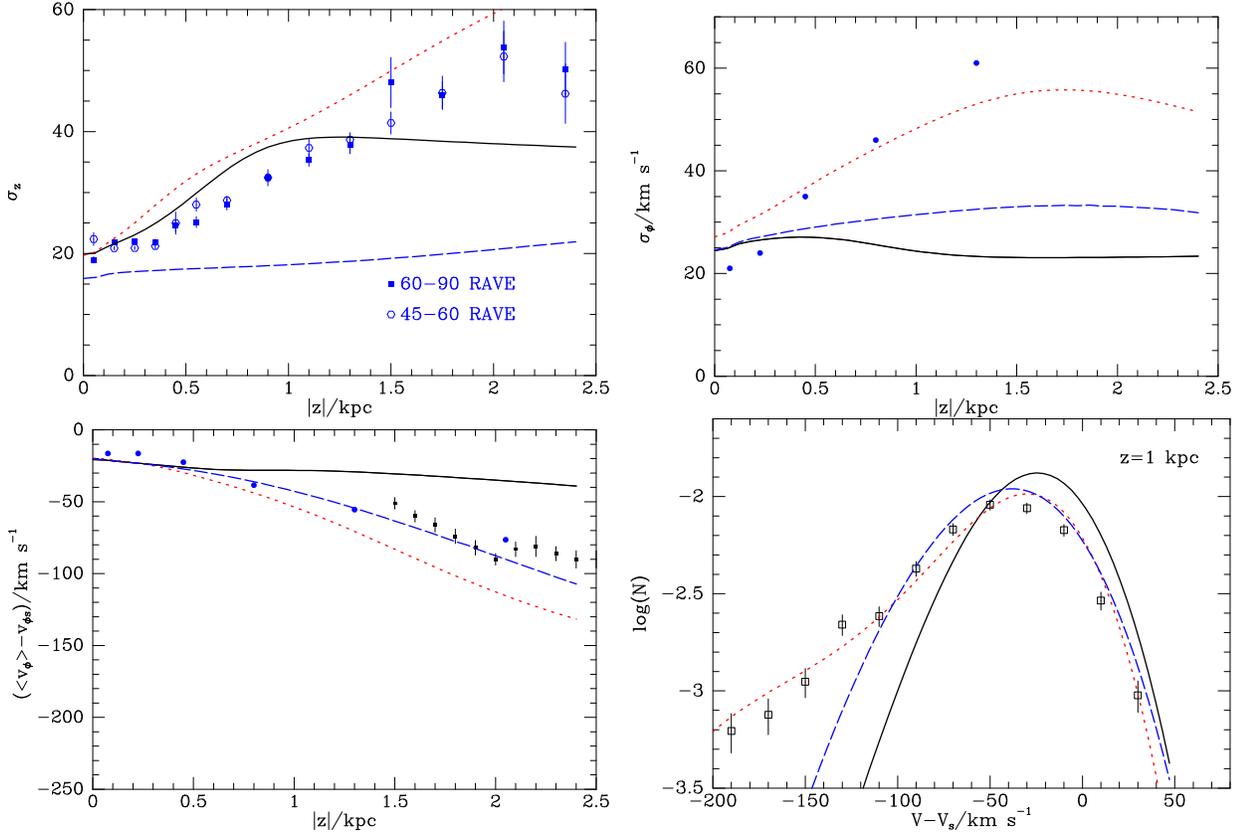

\centerline{
\epsfig{file=c5/PJM12_sigz.ps,width=.45\hsize}\quad
\epsfig{file=c5/PJM12_sp.ps,width=.45\hsize}
}
\centerline{
\epsfig{file=c5/PJM12_vp.ps,width=.45\hsize}\quad
\epsfig{file=c5/PJM12_V1.ps,width=.45\hsize}
}
\caption{The predictions of \df s fitting in Potential II. The coding of the
curves is as in Figs.~\ref{fig:m2eA} and \ref{fig:m2eB}: blue dashed
curve a pure thin disc fitted to only the GCS velocity histograms; full black
curves the result of using a thick disc to obtain a good fit to the Gilmore
\& Reid data for $\rho(z)$; red dotted curves the result of adjusting
all nine parameters of the \df\ simultaneously.}\label{fig:PJM12}
\end{figure*}

The top left panel shows that at both the smallest radii (top) and largest
radii (bottom) the vertical density profile clearly comprises two
straight-line segments, indicative of accurately exponential vertical density
profiles for each disc. The height at which the thick disc becomes dominant
shifts slowly upwards from $\sim0.7\kpc$ at $R=6\kpc$ and the transition
becomes less prominent with increasing radius as the scale-height of the
thick disc decreases with increasing $R$. This decrease reflects the rather
steep decline in $\sigma_{z0}$ implied by the scale-length $R_\d/q=3.24\kpc$.
The scale-height of the thin disc slowly increases with radius.

In the top-right panel a red straight-line has been drawn between points at
$R=6.5$ and $9.75\kpc$, and we see that in the plane the density profile is
accurately exponential. The scale-length of this exponential is
$R_\d=2.64\kpc$, slightly larger than the scale-length of the thin-disc's
\df\ ($2.58\kpc$) and of the thin disc that generates the potential
($2.4\kpc$). As one moves away from the plane, the scale-length is constant
in the region dominated by the thin disc, but at $z\sim500\pc$ it begins to
fall, reaching $1\kpc$ at $z=2.4\kpc$. This behaviour reflects the steep
temperature gradient of the thick disc, which makes the density well above
the plane fall rather slowly with $z$ at small $R$ and steeply with $z$ at
large $R$. 

\cite{Robin03} fitted the 2MASS star counts to a model of the stellar density
that had quite complex functional forms rather than simple double
exponentials for the discs, but their model implies $R_\d\simeq2.5\kpc$ for
both the thin and thick  discs and $z_0\simeq0.8\kpc$ for the thick disc. 
\cite{Juric08} infer from SDSS star counts that the thin disc has scale
lengths $z_0=300\pc$ and $R_\d=2.6\kpc$, while the thick disc has
$z_0=0.9\kpc$ and $R_\d=3.6\kpc$. \cite{Bovy12} by contrast argue that the
disc is a superposition of an infinite number of chemically homogeneous
populations, with each population characterised by values of $z_0$ and $R_\d$
that vary from $(0.2,4.5)\kpc$ at the metal-rich extreme to $(1,2)\kpc$ at the
metal-poor extreme. In particular, these two studies, both based on SDSS star counts,
reach opposite conclusions regarding the ratio of the radial scale lengths of
the thin and thick discs.

The middle  panels of \figref{fig:global} show how the mean-streaming
velocity (left) and $\sigma_\phi$ (right) vary with $z$. Again the red curves
are for $R=6\kpc$. At $|z|<1\kpc$ the decline in $\vpbar$ with  $z$ is
fastest at the smallest radii, but at greater heights $\vpbar$ declines
fastest with $z$ at the largest radii. At a given $z\lta1.5\kpc$, $\sigma_\phi$
is largest at small radii, but this is not true at $z\simeq2\kpc$ because at
large radii $\sigma_\phi$ starts to rise rapidly at $z\simeq1\kpc$.
In general $\sigma_\phi$ mirrors $\vpbar$, rising as $\vpbar$ falls.

The bottom-left panel of \figref{fig:global} shows that at $R=6\kpc$ (top
curve)
$\sigma_z$ rises most rapidly with $z$ for $z\lta0.9\kpc$, while at large $R$
the rise of $\sigma_z$ is gradual below $z\sim0.8\kpc$ and then becomes
rapid.

\subsection{Fits in Potential II}

We now briefly discuss results obtained by fitting \df s in Potential II,
which is characterised by larger values of $R_0$ and the local circular speed
$\Theta_0$. There are two reasons for turning to this potential. First, there
are indications that $R_0>8\kpc$ and $\Theta_0\gta240\kms$
\citep[e.g.][]{McMillanB10}, and second, \figref{fig:m2eB} shows that \df s in
Potential I cannot simultaneously make $\sigma_\phi$ and $\langle
v_\phi\rangle$ as large as the (possibly suspect) data imply, and one might
imagine this failure reflects inappropriate values of $R_0$ and $v_{\rm
c}(R_0)$.  Table~\ref{tab:fitII} gives the parameters of the \df s chosen by
fitting to the the GCS velocity histograms and the Gilmore-Reid density
values in three stages as before, and \figref{fig:PJM12} shows the
corresponding predictions.

\begin{table}
\caption{Parameters of the \df\ chosen by {\it amoeba} for Potential II.
Column (a) shows the thin-disc \df\ chosen to optimise the fits to just the
GCS velocity distributions.  Column (b) gives the parameters obtained when we
add both a thick disc and data for $\rho(z)$. Column (c) shows the \df\
chosen when {\it amoeba} is given the opportunity to adjust all parameters
simultaneously, starting with the \df\ of column (b).}\label{tab:fitII}

\begin{center}
\begin{tabular}{llccc}
\hline
&&(a)&(b)&(c)\\
\hline
Thin&$\sigma_{r0}$&40.9&40.9&42.3\\
&$\sigma_{z0}$    &27.1&27.1&20.9\\
&$R_\d$           &2.29&2.29&3.14\\
&$q$              &0.239&0.239&0.246\\
\hline
Thick&$\sigma_{r0}$&-   &28.2&28.3\\
&$\sigma_{z0}$     &-   &64.7&40.4\\
&$R_\d$            &-   &2.25&3.62\\
&$q$               &-   &0.283&1.070\\
&$F$     &0   &0.395&0.709\\
\hline
$\chi^2$&&19.2&12.2&9.16\\
\hline
\end{tabular}
\end{center}
\end{table}

The second \df\ in the sequence, whose predictions are shown by black full
lines in \figref{fig:PJM12}, is less successful than the corresponding \df\
in Potential I (\figref{fig:m2eB}) because it has too much rotation and too
little random velocity; in Tables \ref{tab:fitI} and \ref{tab:fitII} this
\df\ is stands out for its exceptionally large value of
$\sigma_{z0}=64.7\kms$ for the thick disc. When {\it amoeba\/} is allowed to
adjust all the \df's parameters simultaneously, it increases the scale
lengths of both discs from $\sim2.3\kpc$ to $3.14$ and $3.62\kpc$ for the
thin and thick discs, respectively, and reduces $\sigma_{z0}$ for both discs
to $21$ and $40\kms$, respectively. The predictions of the final \df\ are
shown by the red dotted curves in \figref{fig:PJM12}. They are less
successful than the corresponding predictions in Potential I in that the
values of $\sigma_z$ are too large and the other predictions are only
comparably successful. The excessive values of $\sigma_z$ suggest that
Potential II has a disc that is too massive, and that a larger fraction of
the mass that keeps $\Theta_0$ high at $8.37\kpc$ should reside in the dark halo.

\begin{figure*}
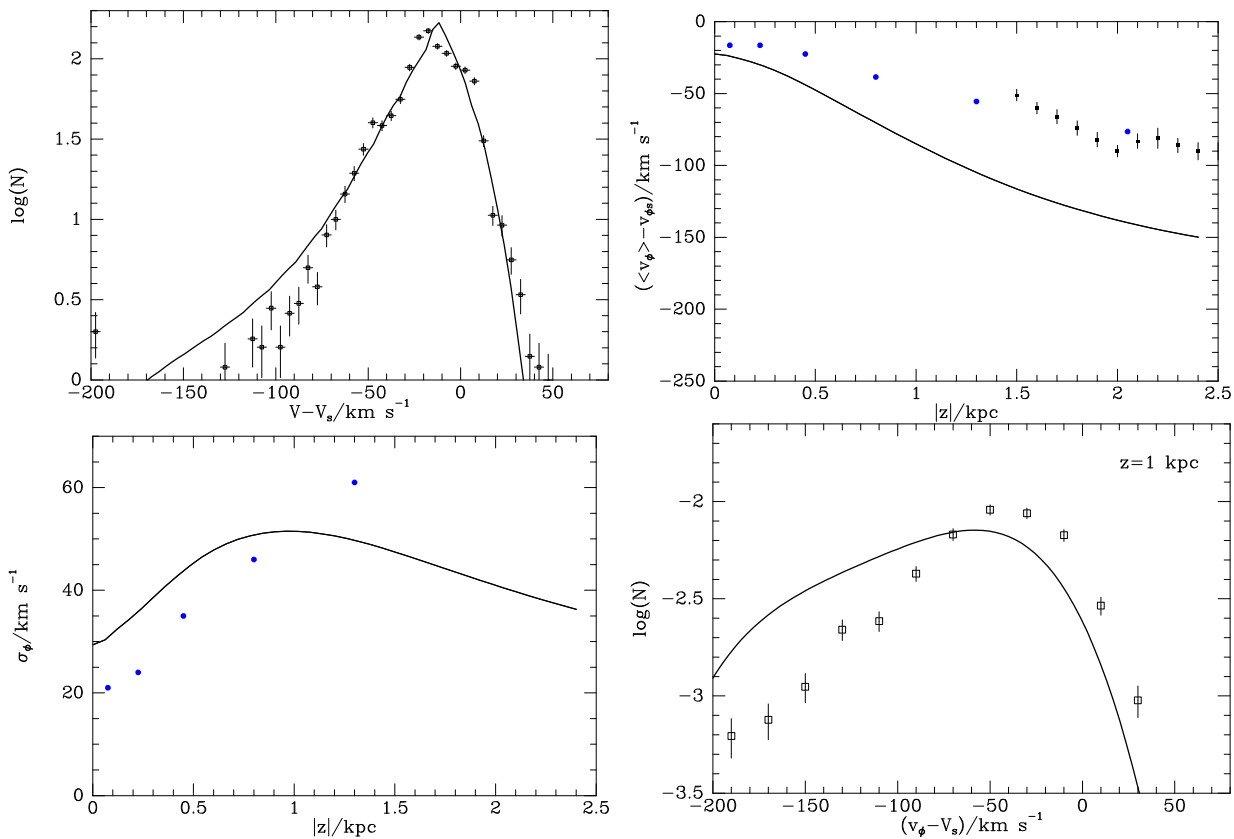

\centerline{\epsfig{file=c5/m2e_V_2.ps,width=.45\hsize}\quad
\epsfig{file=c5/m2e_vp_2.ps,width=.45\hsize}}
\centerline{\epsfig{file=c5/m2e_sp_2.ps,width=.45\hsize}\quad
\epsfig{file=c5/m2e_V1_2.ps,width=.45\hsize}}
 \caption{A model with a hot thick disc  in Potential I. Column (e) of Table
\ref{tab:fitI} lists the parameters of a \df\ with a thick disc that is
radially hotter than the thin disc. This \df\ provides an excellent fits to
the vertical density profile and the distribution of $W$ components of GCS
stars, and a reasonable fit to $\sigma_z(z)$. However, it consistently fails
to reproduce observations of the distribution of $v_\phi$ components because
at large $z$ it places too many stars on highly eccentric orbits.
}\label{fig:m2eB_2}
\end{figure*}

\section{Discussion}\label{sec:discuss}

An aspect of the fitting process that is troubling is that using only a thin
disc {\it amoeba\/} is able to fit the wings as well as the cores of the $U$
and $V$ distributions of local stars -- one would have expected the wings of
these distributions to be filled out by the thick disc, just as is the case
for the $W$ distribution. A consequence of this filling of the wings in $U$
and $V$ by the thin disc is that the  thick discs subsequently fitted have
unexpectedly small radial velocity-dispersion parameters, and these discs
invariably have significantly larger vertical dispersions than radial ones.

\figref{fig:m2eB_2} shows the result of attempting to remedy this situation
by fixing the parameters of the thin-disc \df\ to those listed in column (e)
of Table~\ref{tab:fitI} and then asking {\it amoeba\/} to choose the
thick-disc parameters that minimise the residuals between the model and (i)
the Gilmore-Reid points for $\rho(z)$, (ii) the GCS counts at $|U|>30\kms$ and
(iii) the GCS counts at $|W|>20\kms$. The chosen \df\ provides perfect fits to
$\rho(z)$ and $N(W)$. The fit to $N(U)$ is good at $|U|\gta30\kms$ but
significantly too sharply peaked at $|U|\lta15\kms$. The fit to $\sigma_z(z)$
is excellent at $z\lta1.2\kpc$ but is slightly lower than the data indicate
at greater heights. However, \figref{fig:m2eB_2} shows that this model
predicts too many stars with low $v_\phi$. The surplus of
low-angular-momentum stars becomes more marked as one moves away from the
plane, and is a clear consequence of the thick disc being too hot radially.
This experiment forces us to the conclusion that the thick disc really is
hotter vertically than horizontally, and is indeed radially cooler than the
thin disc. Moreover, it implies that the ability of the thin-disc \df\ to fit
even the wings of the GCS $U$ and $V$ distributions does not arise from an
incorrect choice for the thin-disc's \df's dependence on $J_r$, but
reflects the fact that these wings are populated by stars that do not stray
far from the plane.

When {\it amoeba\/} is permitted to adjust all nine parameters of the
combined \df\ simultaneously, it achieves slightly better representations of
the data for the solar cylinder by adopting \df s that have unexpected, even
implausible, radial structure. In particular there is a systematic tendency
to choose for the thick disc a large radial scale-length and a large (and
compensating) value of the parameter $q$ that controls the radial gradient of
velocity dispersion.  It seems that although data for the solar cylinder do
very strongly constrain the \df s of the individual discs, they do not
suffice to prevent one disc being played off against the other in unphysical
ways. It is likely that such trade-offs would be suppressed if we had data
that spanned a wider radial range.

The ability to distinguish chemically several populations of stars is a
crucial aspect of astronomy that has been neglected in this work. The
division of the disc into thin and thick components acquires objective
meaning only when it is possible to distinguish stars of the two discs by age
or metallicity \citep[e.g.][\S10.4.3]{BinneyM98}. The present models seem to
require that the radial and azimuthal velocity dispersion of the population
of $\alpha$-enhanced (and thus thick-disc) stars is smaller than its vertical
velocity dispersion.  This is a prediction that can be tested when large
samples of photometrically selected stars with known abundances become
available.

Whatever the outcome of this test, each chemically distinguishable population
has an independent \df, and the requirement that different populations
co-exist within a common gravitational potential will surely provide the
strongest constraints on the Galaxy's mass distribution. Consequently, it is
important to extend our formulae for the \df\ to include chemical properties
such as [Fe/H] and [$\alpha$/H]. We hope to present such extensions shortly.

Once one recognises that the Galaxy contains stars that span a range of age
and chemistry, one has to engage with the differing propensities of stars to
be picked up in a given survey. Some surveys select stars kinematically, some
by colour and all select by apparent magnitude, so to predict from a \df\ the
numbers of stars of each species predicted in a given survey, one has to fold
predictions of type presented here through a code such as {\it Galaxia\/}
\citep{Sharma10} that produces number counts from phase-space distributions.
We hope soon to present results obtained in this way.

We do not quote errors on the parameters of our models for two reasons. First
{\it amoeba\/} merely seeks the minimum of a function, and determining the
errors on the nine parameters and their correlations would involve a
computational effort comparable to that involved in locating the minimum.
Second, the formal errors are of little interest because the uncertainties in
the parameters are not determined by the statistical errors, in the data,
which are for the most part small, but by systematics, such as the existence
of substructure that cannot be represented by the models. In fact, the values
of $\chi^2$ per degree of freedom are quite large ($\sim2$) so formally the
models are inconsistent with the data.

Integral-field units now make it possible to map the line-of-sight velocity
distribution and some chemical information across large parts of the images
of external galaxies. Traditionally these data have been interpreted with
either Schwarzschild models \citep{Cappellari07} or models based on the Jeans
equations \citep{JAM}.  These data could be interpreted with models similar
to those presented here with greater ease than is possible with Schwarzschild
models and greater rigour than the Jeans equations allow -- the latter
require an arbitrary closure assumption. This seems a fruitful direction for
future work.

\section{Conclusions}\label{sec:conclude}

The simplest dynamical models of our Galaxy have distribution functions that
are analytic functions of the action integrals of motion. We have fitted such
\df s to measurements of the distribution of stellar velocities in the
immediate neighbourhood of the Sun, and to these data in conjunction with an
estimate of the vertical density profile at the solar circle. We have done
this for two models of the Galaxy's gravitational potential that differ in
their values of $R_0$ and $\Theta_0$.

Using the potential with $R_0=8\kpc$ and $\Theta_0=220\kms$, the model
optimised to fit only the local velocity distribution predicts a vertical
density profile that fits the data below $\sim0.5\kpc$ but falls increasingly
below the data at greater distances from the plane. In fact it provides a
good representation of the thin disc but deviates from the data where the
thick disc is important because the local velocity distributions barely
constrain the thick disc.  When a thick disc is added and used to ensure that
the density profile in the solar cylinder agrees with the measurements of
\cite{GilmoreR}, the model correctly (i) predicts a preliminary estimate of
the run of vertical velocity dispersion with $z$ from the RAVE survey, (ii) fits
two sets of measurements of $\vpbar$ at $z<2.5\kpc$ and (iii) predicts the
distribution of $V$ components of SDSS stars seen at $z\sim1\kpc$. The single
failure of this model is to predict values of $\sigma_\phi$ smaller than
those obtained from preliminary analysis of RAVE data. If the adopted
gravitational potential were significantly in error, it should not be
possible to fit simultaneously the vertical profiles of $\rho$ and
$\sigma_z$, so our findings suggest that the adopted potential is close to
the truth.

When all nine parameters of the \df\ are adjusted to refine the fits to the
local velocity distributions and the vertical density profile, a model is
obtained that predicts much better values of $\sigma_\phi$ at the price of
predicting smaller values of $\vpbar$ at $z\gta1\kpc$ than the raw data
imply. After making allowance for observational error,
the model does provide quite a good fit to the measured distribution of
$v_\phi$
components at $z\simeq1\kpc$.

When the same exercise is conducted with a potential in which $R_0=8.37\kpc$
and $\Theta_0=241\kms$, less satisfactory predictions are obtained. Most
strikingly, in this potential $\sigma_z$ is predicted to be larger than the
RAVE data imply, which suggests that this potential is generated by a disc
that is more massive than the Galaxy's disc.

At radii between $R=6\kpc$ and $R=10\kpc$ the favoured model's vertical
density profile is well approximated by two exponentials, a steep one
associated with the thin disc and a much shallower thick-disc profile.  These
profiles meet at an altitude $\sim0.7\kpc$. In this model the scale-height of
the thin increases only slowly with radius, but that of the thick disc
decreases with radius. Below $z\sim0.9\kpc$ the mean-streaming velocity is
similar at all radii and declines only slowly with increasing $z$, especially
at large $R$. Above
$z=0.9\kpc$ and at larger radii the mean-streaming velocity declines more rapidly
with increasing $z$. A decline in mean-streaming velocity is always matched
by an increase in azimuthal velocity dispersion.

A surprising, but apparently robust, prediction of these models is that, in
contrast to the thin disc, the thick disc is hotter vertically than
horizontally. When kinematically unbiased samples of stars with measured
chemical compositions are available, it will be possible to test this
prediction observationally.

A \df\ of the type used here predicts many observables that we have not
presented -- for example the spatial distribution of stars of a given age or
of the thick-disc stars, or the distributions of $U$ and $W$ components of
velocity at $z\sim1\kpc$ or any other altitude. We will release programs that
calculate these predictions and it will be instructive to compare the
predictions with further observations.

\section*{Acknowledgements}
I thank P.J. McMillan for providing the parameters of Potential II and for
comments on an early version of the paper.

\section*{Appendix: Multiple integrals}\label{sec:ints}

Evaluation of a model's observables, such as the density
$\rho=\int\d^3\vv\,f(\vJ)$ and velocity moments
\[
\sigma_{ij}^2={1\over\rho}\int\d^3\vv\,v_iv_jf(\vJ)
\] 
 from its distribution function (\df) $f(\vJ)$ involves many multiple
integrals and it is important to do these efficiently. We do
three-dimensional integrals with the aid of an oct-tree: the integral over a
cubic region of volume $V$ is first estimated from values of the integrand $f$ at the
corners and the centre of the cube as
 \[
I=\fracj12V[f(\hbox{centre})+\fracj18\sum f(\hbox{corners})].
\]
 Then the integral is evaluated from the same formula applied to each of the
eight sub-cubes into which the parent cube can be divided. If the sum of the
sub-integrals differs by a given amount from the first estimate and the
side-length of the sub-cubes exceeds a given length, the algorithm is run
recursively on each of the sub-cubes. If either of these conditions is
violated, the sum of the estimates for the sub-cubes is accepted as the value
of the integral over the parent cube. For increased efficiency integrands for
several moments of the \df\ are evaluated simultaneously, with the criteria
for exit based exclusively on the lowest moment. Analogous recursive
algorithms are used to estimate one- and two-dimensional integrals.

\label{lastpage}

\end{document}